\documentclass{elsart}

\usepackage{graphicx}
\usepackage{epsfig,array}
\usepackage{booktabs}

\newcommand{\beq}{\begin{equation}}
\newcommand{\eeq}{\end{equation}}
\newcommand{\beqa}{\begin{eqnarray}}
\newcommand{\eeqa}{\end{eqnarray}}
 
\newcommand\eq[1]{eq.~(\ref{eq:#1})} 
\newcommand\tab[1]{{\footnotesize {\bf Table}~[{\bf\ref{#1}}]}} 
\newcommand\sect[1]{sec.~\ref{sec:#1}}

\newcommand{\sla}[1]%
        {\kern .25em\raise.18ex\hbox{$/$}\kern-.75em #1}
\newcommand{\mybar}[1]%
        {\kern 0.8pt\overline{\kern -0.8pt#1\kern -0.8pt}\kern 0.8pt}

\newcommand\mycaption[1]{\caption{\footnotesize \sf #1}} 

\begin{document} 

\begin{frontmatter}

\title{Heavy--light decay constants in the continuum limit of quenched lattice QCD}

\author[romeII]{G. M. de Divitiis}
\author[romeII]{M. Guagnelli}
\author[fermi]{F. Palombi}
\author[romeII]{R. Petronzio}
\author[romeII]{N. Tantalo}

\address[romeII]{University of Rome ``Tor Vergata'' and INFN sez. RomaII, 
	Via della Ricerca Scientifica 1, 
	I-00133 Rome}

\address[fermi]{Enrico Fermi Research Center, 
	Via Panisperna 89a, I-00184 Rome}

\begin{abstract}
We compute the decay constants for the heavy--light pseudoscalar mesons in the quenched approximation and 
continuum limit of lattice QCD. Within the Schr\"odinger Functional framework, we make use of the step scaling method, 
which has been previously introduced in order to deal with the two scale problem represented by the coexistence of a 
light and a heavy quark. The continuum extrapolation gives us a value $f_{B_s} = 192(6)(4)$ MeV for the $B_s$ meson decay constant 
and $f_{D_s} = 240(5)(5)$ MeV for the $D_s$ meson.
\end{abstract}


\end{frontmatter}

\section{Introduction}
\label{sec:1}

The amount of CP violation occurring in the in the Standard Model
depends upon the unitary Cabibbo-Kobayashi-Maskawa (CKM) matrix. 
This is explored through the so called Unitarity Triangle Analysis
\cite{Ciuchini:2000de,Hocker:2001xe,Buras:2002yj} that
requires a deep theoretical understanding of the $B$ mesons decay properties,
given the increasing accuracy of the $B$-factory experiments (for
a recent review see \cite{Stocchi:2002yi}).
A crucial quantity, the $B$ meson decay constant $f_B$, has already been calculated
by means of different techniques: Heavy Quark Effective Theory (HQET)
\cite{Broadhurst:1992fc,Bagan:1992sg,Neubert:1992sp} QCD Sum Rules 
\cite{Aliev:1983ra,Dominguez:1987ea,Narison:1987qc,Reinders:1988vz,Colangelo:1991ug,Penin:2001ux,Jamin:2001fw} 
and lattice QCD (LQCD).\\

The latter moves from first principles and has to face the problem of 
properly accounting two largely separated energy scales, i.e. the heavy and light quark masses.
This imposes stringent limits on the values of the lattice bare parameters:  
the lattice spacing has to be small enough to allow a good description
of the highly localized heavy quark, and the number of lattice points has to be large enough 
to accommodate the widely spread light quark.
A direct calculation would require lattice sizes of the order $100^4$ points. 
These limits are too much demanding for present supercomputers, even in the quenched approximation, and
in the case of full QCD also for the next generation supercomputers.
To overcome these difficulties, different strategies have been adopted
(see \cite{Yamada:2002wh,Becirevic:2002zp} for recent reviews).\\

One way consists in working with propagating heavy quarks with masses
in the region of the physical charm quark and to extrapolate the results
to the $b$--quark, using HQET scaling laws 
\cite{Becirevic:1998ua,Bernard:1998xi,AliKhan:2000eg,Bowler:2000xw,Lellouch:2000tw,Bernard:2002pc}.
The systematic errors in these calculations are dominated by the uncertainty on the functional
form to be used in the extrapolation, depending in turn upon the limits of
validity of HQET when applied to the $c$--quark.
Another method comes from the non--perturbative matching
of lattice QCD and HQET in the static limit \cite{Kurth:2000ki,Heitger:2003xg}.
Other possibilities are offered by the Non Relativistic approximation (NRQCD),
where the heavy quark mass expansion is taken in the operators as well as in
the Lagrangian \cite{Ishikawa:1999xu,AliKhan:2001jg}, and by the Fermilab approach \cite{El-Khadra:1998hq}, where
one expands in  the mass of the propagating heavy quark in lattice units.
All these methods give results with systematic errors that makes them
fully compatible within the error bars \cite{Yamada:2002wh,Becirevic:2002zp}.\\

In a previous publication \cite{Guagnelli:2002jd} we have introduced a new method to perform
the determination of $f_B$ and, more generally, to face two scales problems in lattice QCD.
The so called \emph{step scaling method} \cite{Yamada:2002wh} has been
applied to give a first numerical result for this quantity at fixed lattice spacing
and, in \cite{deDivitiis:2003iy}, to perform the first calculation of
the $b$--quark mass in the continuum limit of quenched lattice QCD. 
The idea behind the method consists in using a QCD propagating $b$--quark on a small
volume, calculating the \emph{finite volume effects} on the heavy--light decay constants and, finally, 
using the very mild dependence of these effects upon the heavy quark mass,  
to obtain a final result in a large volume.\\

In this paper we extrapolate our results to the continuum by repeating the
various steps of the previous calculation at different
lattice spacings and fixed physical quantities. The major assumption undergoing our method, i.e.  
that the finite size effects on an heavy--light observable have a milder
dependence upon the heavy quark mass than the quantity itself, 
is shown to hold in the continuum limit.

The general features of the method are described in \sect{method}, in \sect{regren} we set
the notations, the numerical results are presented in \sect{numsim}. Conclusions are drawn
in \sect{concl}. 

\section{Overview of the method}
\label{sec:method}

A detailed explanation of the method can be found in \cite{deDivitiis:2003iy}; here we shortly
review  the basic features in order to set the notations.

A non--perturbative determination of the heavy--light decay constants from lattice QCD should take into account the masses of a 
heavy and a light propagating quark. The step scaling method faces the challenging 
requirements discussed above by adopting a two--step strategy. 
As a first step, the decay constant is computed on a small volume, where the light quark
is squeezed, and the heavy quark propagates with a high resolution. 
At this stage, the mass of the heavy quark can be raised up to very large values, 
with controlled  discretization errors, and the decay constant can be directly simulated 
at the physical heavy quark mass. As a second step, 
the finite size effects of this calculation are removed by evolving 
the decay constant toward large volumes. The evolution is realized according to the identity  
\beq 
f_{h\ell}(L_\infty) = f_{h\ell}(L_0)\ \frac{f_{h\ell}(L_1)}{f_{h\ell}(L_0)}\ 
\frac{f_{h\ell}(L_2)}{f_{h\ell}(L_1)}\cdots, \qquad L_0 < L_1 < L_2 \dots
\label{eq:starting}\eeq
where the basic ingredient is the ratio of the decay constants computed on 
two different volumes at the same values of the mass parameters
\beq
\sigma( m_\ell, m_h, L_{k-1} ) = \frac{f_{h\ell}(m_\ell, m_h, L_k)}{f_{h\ell}(m_\ell, m_h, L_{k-1})}\biggr|_{L_k \ = \ sL_{k-1}}
\label{eq:ssf}\eeq
Throughout the paper we refer 
to the step scaling function in the continuum limit as $\sigma$ (greek lowercase) 
and  to the step scaling function at finite lattice spacing as $\Sigma$ (greek uppercase). 

This quantity represents a non--perturbative calculation of the finite volume effects. 
In principle, its dependence on the quark masses can be very 
different from the one of the decay constants themselves. 
In effects, it has been shown \cite{Guagnelli:2002jd} that the $\sigma$--ratio's are 
characterized by a very slight linear dependence upon the inverse of $m_h$, 
due to cancellations of additional heavy quark mass dependences between the 
numerator and the denominator of \eq{ssf}. This suggests a concrete way to connect the finite volume decay constant to physical volumes:
\begin{itemize}

\item given a couple of physical volumes $(L_{k-1},L_k)$ and a finite lattice spacing $a$, the step scaling function is simulated on 
      the lattice for a set of heavy and light quark masses. 
      In order to identify the quark mass on a finite volume, a RGI quark mass scheme is adopted 
      \cite{Gasser:1985gg,Capitani:1998mw} and units are fixed through the $r_0$ scale \cite{Guagnelli:1998ud,Necco:2001xg}.
      Throughout the paper we fix $r_0 = 0.5$ fm. 
      The light quark masses are kept around the strange mass throughout the whole procedure.

\item A set of different simulations are done at fixed physical volumes $(L_{k-1},L_k)$ but with different lattice spacings,
      in order to perform the continuum extrapolation of the step scaling function at given heavy and light RGI quark masses.
      The ratio $s$ between the two volumes should be chosen small enough to cope with the increase of lattice sites
      without exceeding computational resources. On the other hand, it should be large enough to reach large volumes in few
      steps. A value $s=2$ is a good compromise.
      The continuum step scaling functions are then linearly extrapolated in the 
      inverse of the RGI heavy quark masses up to $m^{RGI}_c$ or $m^{RGI}_b$, according to the heavy flavors of the meson.
     
\item As a starting value for the finite volume, we chose to set $L_0 = 0.4$ fm. This allows to reach a volume $L_2=1.6$ fm, 
      after just two evolution steps, which is adequate to accommodate the heavy--light mesons at the physical
      values of the light quark masses. 

\end{itemize}

In order to match subsequent steps, the knowledge of the bare coupling $g_0(a)$ 
as a function of the lattice spacing is required at very small couplings.
The problem has been recently addressed in \cite{Guagnelli:2002ia} and solved by a renormalization group analysis.

\section{Observables}
\label{sec:regren}

The step scaling function is calculated within the SF
\cite{Luscher:1992an,Sint:1994un}, which has already been applied 
to a number of different 
finite size problems 
\cite{Luscher:1994gh,Capitani:1998mw,Bode:2001jv,Guagnelli:2003hw,Heitger:2003xg}. 
The lattice topology is $T\times L^3$ with 
periodic boundary conditions on the space directions and Dirichlet 
boundary conditions along time. We use the following set of parameters
\beq
T=2L, \qquad C = C' = 0, \qquad \theta = 0
\eeq
where $C$ and $C'$ represent the boundary gauge fields and $\theta$ is a topological phase which affects the periodicity 
of the fermion in the space directions. 
Lattice discretization is performed using non--perturbative $O(a)$ improved clover 
action  \cite{Luscher:1997ug} and operators. In order to set the notation, let 
\beqa
A_\mu(x) &=& \overline{\psi}_i(x) \gamma_\mu \gamma_5\psi_j(x) 
\nonumber \\ \nonumber \\
P(x) &=& \overline{\psi}_i(x) \gamma_5 \psi_j(x) 
\eeqa
be the axial current and the pseudoscalar density 
($i$ and $j$ are flavor indices).
The improvement of the axial current is obtained through the relations
\beq
A^I_\mu(x) = A_\mu(x) + a c_A \ \tilde{\partial}_\mu P(x)
\eeq
where $\tilde{\partial}_\mu = (\partial_\mu + \partial_\mu^*)/2$ and $\partial_\mu$, $\partial^*_\mu$ are the usual 
forward and backward lattice derivatives respectively.
For what concerns the improvement coefficients $c_A$, we use the non--perturbative results of \cite{Luscher:1997ug}. 
The correlation functions used
to compute the meson decay constants are defined by probing the previous operators with appropriate boundary quark sources  
\beqa
\mathcal{F}^I_A(x_0) &=& -\frac{a^6}{2} \sum_{\bf y,z}\langle \overline{\zeta}_j({\bf y}) 
\gamma_5 \zeta_i({\bf z}) A^I_0(x) \rangle
\nonumber \\ \nonumber \\
\mathcal{F}_P(x_0) &=& -\frac{a^6}{2} \sum_{\bf y,z}\langle \overline{\zeta}_j({\bf y}) 
\gamma_5 \zeta_i({\bf z}) P(x) \rangle 
\label{eq:correlations}
\eeqa
where $\zeta_i({\bf y})$ and $\overline{\zeta}_i({\bf y})$ can be considered as quark and anti--quark boundary 
states.

The renormalization of the axial current is realized according to the following relation
\beq
A^R_\mu(x_0) = Z_A \ (1 + b_A \ am ) \ A^I_\mu(x_0) 
\label{eq:rencorr}
\eeq
here $am$ is the bare quark mass defined as
\beq
am_i = \frac{1}{2} \left[\frac{1}{k_i} - \frac{1}{k_c} \right] 
\label{eq:barequarkmass}
\eeq
The renormalization constant $Z_A$ has 
been computed non perturbatively in \cite{Luscher:1997jn}.
For the improvement coefficient $b_A$ we use the perturbative results quoted in \cite{Sint:1997jx}
(at the values of the bare coupling, $\beta \simeq 7.0$, used in the numerical simulations 
the one--loop contribution to $b_A$ differs from the tree--level of $10$\%).\\

The so--called bare current quark masses are defined through the lattice version of the PCAC relation
\beq
m^{WI}_{ij} = \frac{ \tilde{\partial_0}\mathcal{F}_A(x_0) + a c_A \partial_0^* \partial_0 \mathcal{F}_P(x_0)  }{2 \mathcal{F}_P(x_0)} 
\label{eq:nondiagonal}
\eeq
These masses are connected to the renormalization group invariant (RGI) quark masses, according to the definitions
given in \cite{Gasser:1985gg}, through a renormalization factor
which has been computed non--perturbatively in \cite{Capitani:1998mw}:
\beq
m_{ij}^{RGI} = Z_M(g_0) \ \left[ 1 + (b_A-b_P)\ \frac{am_i+am_j}{2} \right] \ m^{WI}_{ij}(g_0)
\label{eq:rgimassij}
\eeq
where $am_i$ is defined in eq.~(\ref{eq:barequarkmass}).

The combination $b_A-b_P$ of the improvement coefficients of the axial current and pseudoscalar density has been
non--perturbatively computed in \cite{deDivitiis:1998ka,Guagnelli:2000jw}. The factor $Z_M(g_0)$ is known with very high precision in a 
range of inverse bare couplings that does not cover all the values of $\beta$ used in our simulations.
We have used the results reported in table~(6) of ref.~\cite{Capitani:1998mw} 
to parametrize $Z_M(g_0)$  in the enlarged range of $\beta$ values $\left(5.9,7.6\right)$.

The RGI mass of a given quark is obtained from eq.~(\ref{eq:rgimassij}) using the diagonal
correlations
\beq
m^{RGI}_i = m_{ii}^{RGI} 
\label{eq:rgimass1}
\eeq
From non--diagonal correlations in eq.~(\ref{eq:rgimassij}) 
one obtains different $O(a)$ improved definitions of the RGI $i$--quark mass 
for different choices of the $j$--flavor:
\beq
m^{RGI}_{i_{\{j\}}} = 2 m^{RGI}_{ij} - m^{RGI}_{jj} 
\label{eq:rgimassi}
\eeq
All these definitions must have the same continuum limit because the dependence upon the $j$--flavor 
is only a lattice artifact.
Further, for each definition we use in eq.~(\ref{eq:nondiagonal}) either standard lattice
time derivatives as well as improved ones \cite{deDivitiis:1998ka,Guagnelli:2000jw}.

Another non--perturbative $O(a)$ improved definition of the RGI quark masses can be obtained starting from the
bare quark mass 
\beq
\hat{m}_i^{RGI} = Z_M(g_0) \ Z(g_0) \ \left[ 1 + b_m\ am_i \right] \ m_i 
\label{eq:rgimass2}
\eeq
where the improvement coefficient $b_m$  
and the renormalization constant 
\beq
Z(g_0) = \frac{Z_m Z_P}{Z_A} 
\eeq 
have been non-perturbatively computed in ref.~\cite{deDivitiis:1998ka,Guagnelli:2000jw}.

\begin{figure}[t]
\begin{center}
\epsfig{file=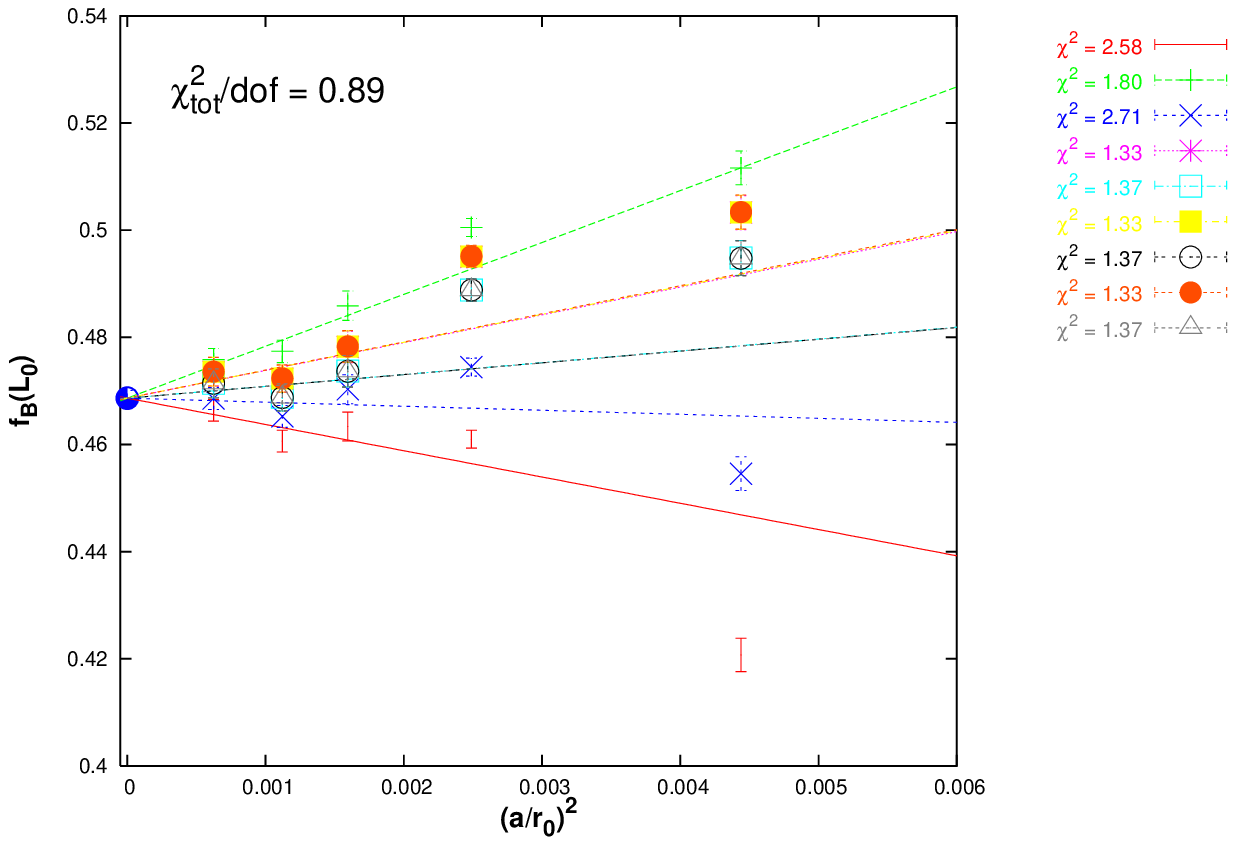,width=12cm}
\mycaption{Continuum extrapolation on the small volume of $f_{B_s}(L_0)$.
The different values of the decay constant correspond to different definitions
of the RGI quark masses given in equations
(\ref{eq:rgimass1}), (\ref{eq:rgimassi}) and (\ref{eq:rgimass2}). 
For each definition the two points at largest lattice spacing
has been shown but not included in the fit. Units are in GeV.
\label{fig:A-ContSV}}
\end{center}
\end{figure}

Equations (\ref{eq:rgimass1}), (\ref{eq:rgimassi}) and (\ref{eq:rgimass2}) give us different possibilities 
to identify the valence quarks inside a given meson (fixed by the values of the bare quark masses).
The procedure is well defined on small volumes because the
RGI quark mass is a physical quantity that does not depend upon the scale,
given in the SF scheme by the volume, and is defined in terms of local correlations
that do not suffer finite volume effects.
Each pair $\left(m^{RGI}_i,m^{RGI}_j\right)$ fixed a priori is matched, changing
the values of the hopping parameters, 
by the different definitions of equations (\ref{eq:rgimass1}), (\ref{eq:rgimassi}) and (\ref{eq:rgimass2}),
and leads to values of the corresponding decay constants differing by $O(a^2)$ lattice artifacts.
We take advantage of this plethora of definitions by constraining  
in a single fit the continuum extrapolations
(see {\footnotesize {\bf Figure} [{\bf \ref{fig:A-ContSV},\ref{fig:A-S1-Cont},\ref{fig:A-S2-Cont}}]}). \\

The meson masses are extracted from the so--called \emph{effective mass}
\beq
a M_X(x_0) = \frac{1}{2} \ln \left[ \mathcal{F}_X(x_0 - a) / \mathcal{F}_X(x_0 + a)\right]   
\label{eq:effmass}
\eeq
where $\mathcal{F}_X$ is one of the correlations defined in (\ref{eq:correlations}).
Correspondingly, the meson decay constants are defined as
\beq
f_{h\ell} = \frac{2}{\sqrt{L^3 M_A(T/2)}}\frac{\mathcal{F}_A^R(T/2)}{\sqrt{\mathcal{F}_1}}
\eeq
where $\mathcal{F}_1$ is the boundary--to--boundary correlation needed in order
to cancel, in the ratio, the renormalization constants of the
boundary quark fields:
\beq
\mathcal{F}_1 = -\frac{a^{12}}{3L^6} 
\sum_{\bf y,z,u,w}{
\langle \
\overline{\zeta}_j({\bf y}) \gamma_5 \zeta_i({\bf z}) \;
\overline{\zeta'}_i({\bf u}) \gamma_5 \zeta'_j({\bf w})
\ \rangle}
\label{eq:f1}
\eeq 
We want to stress that our choice of defining the decay constant
in the middle of the lattice, $x_0=T/2$, does not introduces other
length scales than $L$ into the calculation.
Indeed we take $T = 2L$ and 
the step scaling technique (see eq.(\ref{eq:starting})) 
connects $x_0 = L_{min}$, where the decay constant has been defined on the smallest volume,
with $x_0 = L_{max}$, where one expects to be free from finite volume effects.\\

\section{Numerical simulations}
\label{sec:numsim}

\noindent In this section we report, step by step, the results of the
calculations.

\begin{table}[t]
\begin{center}
\scriptsize
\begin{tabular}{cccccc}
\midrule
$\beta$ & $L_0/a$         & $k_c$         & $k$      & $\qquad$ $m^{RGI}$ (GeV) $\qquad$ \\
\toprule
      &                 &               &  0.115528 &    7.51(9)   \\
      &                 &               &  0.116762 &    6.91(8)   \\
      &                 &               &  0.123555 &    4.019(43) \\
6.737 & $12$            & 0.13520(1)    &  0.130384 &    1.702(20) \\
      &                 &               &  0.130089 &    1.604(18) \\
      &                 &               &  0.134801 &    0.1347(31)\\
      &                 &               &  0.134925 &    0.0929(29) \\
      &			&	        &  0.135048 &    0.0513(29) \\
\midrule
      &                 &               &  0.120081 &    7.14(8)    \\
      &                 &               &  0.120988 &    6.63(7)    \\
      &                 &               &  0.126050 &    4.024(44)  \\
6.963 & $16$            & 0.134827(6)   &  0.131082 &    1.696(19)  \\
      &                 &               &  0.131314 &    1.591(18)  \\
      &                 &               &  0.134526 &    0.1381(30) \\
      &                 &               &  0.134614 &    0.0978(28) \\
      &			&	        &  0.134702 &    0.0574(28) \\
\midrule			             
      &                 &               &  0.122666 &    7.03(11)   \\
      &                 &               &  0.123437 &    6.53(10)   \\
      &                 &               &  0.127605 &    3.97(6)    \\
7.151 & $20$            & 0.134492(5)   &  0.131511 &    1.716(27)  \\
      &                 &               &  0.131686 &    1.617(25)  \\
      &                 &               &  0.134277 &    0.1257(36) \\
      &                 &               &  0.134350 &    0.0829(33) \\
      &                 &               &  0.134422 &    0.0407(32) \\
\midrule			             
      &                 &               &  0.124176 &    7.11(8)    \\
      &                 &               &  0.124844 &    6.61(20)   \\
      &                 &               &  0.128440 &    4.018(44)  \\
7.300 & $24$            & 0.134235(3)   &  0.131800 &    1.695(19)  \\
      &                 &               &  0.131950 &    1.592(18)  \\
      &                 &               &  0.134041 &    0.1374(27) \\
      &                 &               &  0.134098 &    0.0971(24) \\
      &                 &               &  0.134155 &    0.0567(24) \\
\midrule							             
      &                 &               &  0.126352 &    7.10(8)    \\
      &                 &               &  0.126866 &    6.60(7)    \\
      &                 &               &  0.129585 &    4.016(44)  \\
7.548 & $32$            & 0.133838(2)   &  0.132053 &    1.698(19)  \\
      &                 &               &  0.132162 &    1.595(18)  \\
      &                 &               &  0.133690 &    0.1422(27) \\
      &                 &               &  0.133732 &    0.1021(25) \\
      &                 &               &  0.133773 &    0.0618(23) \\
\bottomrule
\end{tabular}
\mycaption{Simulation parameters  at $L_0 = 0.4$ fm. The RGI quark masses
are obtained using eq.~(\ref{eq:rgimass1}).}
\label{tab:SVSimPar}
\end{center}
\end{table}

\subsection{Small volume decay constants ($L_0 = 0.4$ {\rm fm})}
\label{sec:SV}

Simulations of the decay constants on the smallest volume ($L_0 = 0.4$ fm) have been performed at five different lattice spacings
using the geometries $24\times 12^3$, $32\times 16^3$, $40\times 20^3$, $48\times 24^3$ and $64\times 32^3$. 
For each discretization, a set of eight quark masses have been simulated.
Two of the heavy masses have been chosen around the bottom quark $m^{RGI}_b = 6.73(16)$ GeV \cite{deDivitiis:2003iy}. 
Other two have been chosen in the region of the  charm quark $m_c^{RGI}=1.681(36)$ \cite{deDivitiis:2003iy}. 
An additional heavy quark has been simulated with mass $4.00$ GeV.
Three light quark have been simulated with RGI masses of $0.14$ GeV, $0.10$ GeV and
$0.06$ GeV. Using  the accurate determination of the RGI strange quark mass given in \cite{Garden:1999fg}
we have fixed one of the simulated light quarks to be the physical $s$.
We will combine this finite volume calculations with the ones of the step scaling functions
to provide results for the heavy--light decay constants with light quarks around
the the strange in the continuum and on the large volume.
All the parameters of the five different simulations are summarized in \tab{tab:SVSimPar}.

We have obtained different set of data
by using the different definitions of the RGI quark masses
given in the equations (\ref{eq:rgimass1}), (\ref{eq:rgimassi}) and (\ref{eq:rgimass2}). 
The continuum results are thus obtained trough a combined fit of all the set of data,  linear in $(a / r_0 )^2$, 
as shown in {\footnotesize {\bf Figure} [{\bf \ref{fig:A-ContSV}}]} in the case of the $\overline{b}s$ meson.
For each set of data we have included in the fit the three points nearest to the continuum
obtaining a global $\chi^2/dof = 0.89$ to be compared with the $\chi^2$s of each individual definition
listed in the figure.
At this small volume, we are in a region of small bare couplings ($g_0^2\sim 0.85$)
where it is legitimate to use perturbative values for the improvement coefficient $b_A$.
The systematics introduced in the calculation by the continuum extrapolations have
been estimated repeating the fits linear in $(a / r_0 )^2$ including, for each set of data, only
the two points nearest to the continuum.
We find a deviation leading to a corresponding systematic
error of the order of $1$\% that will be given to the results on the large volume added in quadrature 
with an error of about $2$\% coming from the uncertainties on the lattice spacing
and on the renormalization factors.
The latter have been evaluated by 
moving the points as a consequence of the change, within the errors, of
the lattice spacings and of the renormalization constants
and by repeating the whole analysis.

\noindent The numbers we obtain are:
\beq
f_{B_s}(L_0) = 475(2) MeV \qquad f_{D_s}(L_0) = 644(3) MeV
\eeq
The errors quoted at this stage are statistical only,
evaluated by a jackknife procedure. \\
Due to the compression of the low energy scale,
these results are higher than
the large volume ones
obtained after the step scaling functions multiplication chain
(see eqs.~(\ref{eq:fin1}), ~(\ref{eq:fbc}) and ~(\ref{eq:fin2})).

\subsection{First evolution step ($L_0 \to L_1$)}
\label{sec:S1}


\begin{table}[t]
\begin{center}
\scriptsize
\begin{tabular}{cccccc}
\midrule
$\beta$ & $L_0/a$         & $k_c$       & $k$      & $\qquad$ $m^{RGI}$ (GeV) $\qquad$\\
\toprule
      &                 &               & 0.120674 &  3.543(39)  \\
      &                 &               & 0.122220 &  3.114(34)  \\
      &                 &               & 0.126937 &  1.927(21)  \\
6.420 & $8$             & 0.135703(9)   & 0.134304 &  0.3007(36) \\
      &                 &               & 0.134770 &  0.2003(28) \\
      &                 &               & 0.135221 &  0.1028(21) \\
\midrule
      &                 &               & 0.1249   &  3.542(39)  \\
      &                 &               & 0.1260   &  3.136(34)  \\
      &                 &               & 0.1293   &  1.979(22)  \\
6.737 & $12$            & 0.135235(5)   & 0.1343   &  0.3127(38) \\
      &                 &               & 0.1346   &  0.2090(28) \\
      &                 &               & 0.1349   &  0.1080(21) \\
\midrule
      &                 &               & 0.127074 &  3.549(39)  \\
      &                 &               & 0.127913 &  3.153(35)  \\
      &                 &               & 0.130409 &  2.003(22)  \\
6.963 & $16$            &  0.134832(4)  & 0.134145 &  0.3134(38) \\
      &                 &               & 0.134369 &  0.2112(28) \\
      &                 &               & 0.134593 &  0.1086(20) \\
\bottomrule
\end{tabular}
\mycaption{Simulation parameters for the first evolution step $L_0 \to L_1 = 0.8$ fm.
The RGI quark masses are obtained using eq.~(\ref{eq:rgimass1}).}
\label{tab:S1simpar}
\end{center}
\end{table}

\begin{figure}[t]
\begin{center}
\epsfig{file=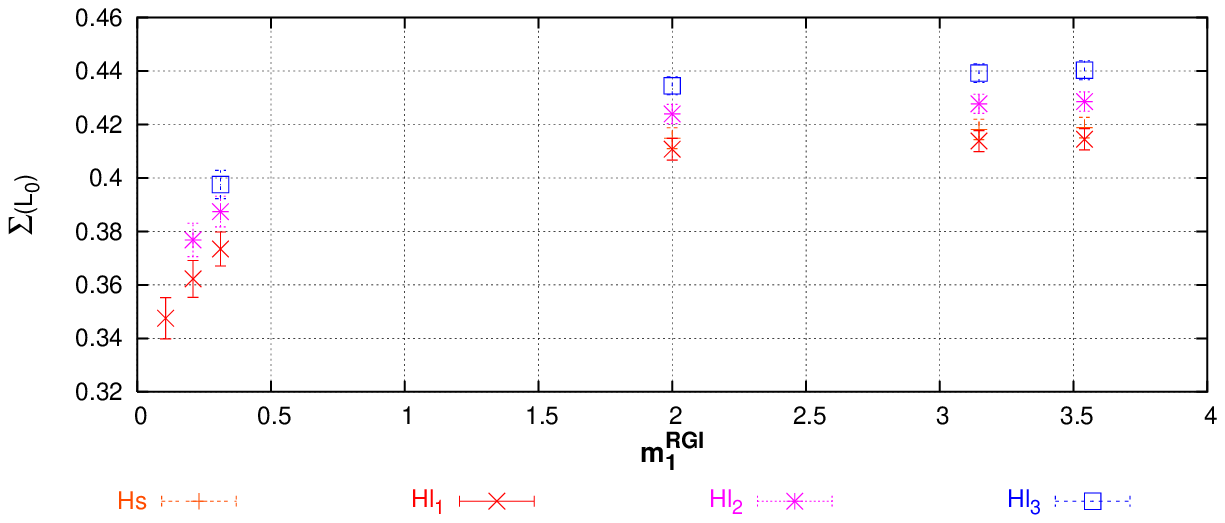,width=12cm}
\mycaption{
The figure shows the step scaling functions $\Sigma(L_0)$ as functions of $m_1^{RGI}$,
for the simulation of the first evolution step
corresponding to $\beta = 6.963$.
The different sets of data correspond to the values of $m_2^{RGI}$.
As can be seen
the step scaling functions approach a plateau for high values of $m_1^{RGI}$. 
Similar plots can be obtained
for the other values of the bare couplings.}
\label{fig:A-S1-sigmas}
\end{center}
\end{figure}

\begin{figure}[t]
\begin{center}
\epsfig{file=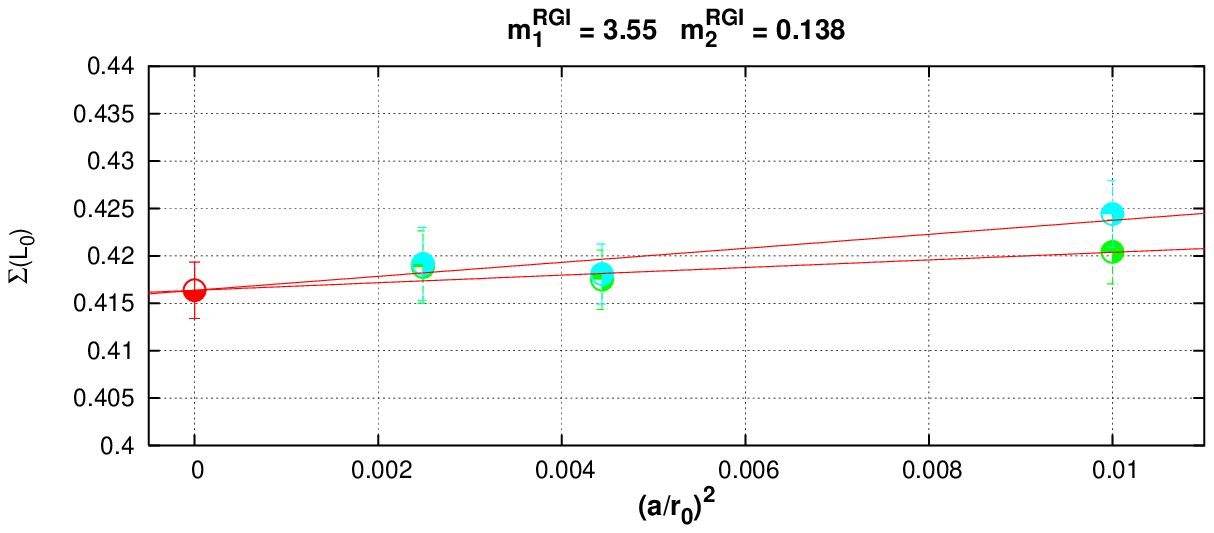,width=12cm}
\mycaption{Continuum extrapolation on the first evolution step, $L_0\mapsto L_1$, of the 
step scaling function, $\Sigma(L_0)$,
of the pseudoscalar meson
corresponding to the heavy quark of mass $m_1^{RGI} = 3.55$ GeV
and to the $s$--quark.
The two sets of data are obtained using the two definitions of RGI quark masses of equations
(\ref{eq:rgimass1}) and (\ref{eq:rgimass2}). 
Units are in GeV.
Similar plots can be obtained for the other 
combinations of quark masses used in our simulations.
\label{fig:A-S1-Cont}}
\end{center}
\end{figure}

\begin{figure}[t]
\begin{center}
\epsfig{file=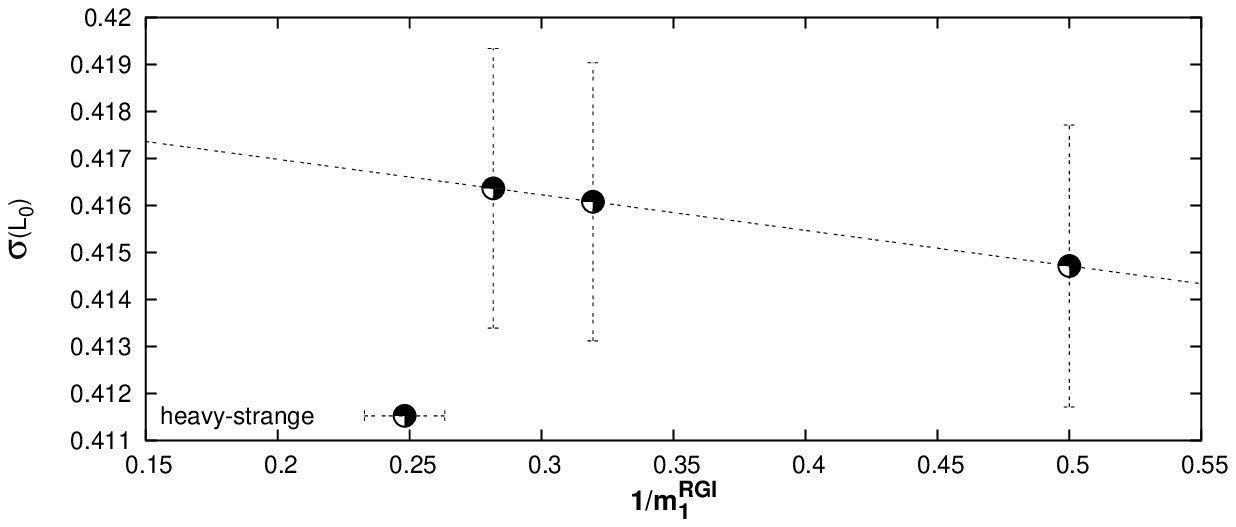,width=12cm}
\mycaption{
The figure shows the continuum extrapolated step scaling functions $\sigma(L_0)$
as functions of $1/m_1^{RGI}$.
The heavy extrapolations are shown only for the $\overline{h}s$ set of data.}
\label{fig:A-S1-sigmas2}
\end{center}
\end{figure}

The finite volume effects on the decay constants calculated on $L_0$,
are measured by doubling the volume, $L_1 = 0.8$ fm, and by using the step scaling
function of eq~(\ref{eq:ssf}).

The continuum extrapolations have been obtained by simulating the step scaling functions with
three different discretizations  of $L_0$, i.e $16\times 8^3$, $24\times 12^3$ and $32\times 16^3$. 
The volume $L_1$ has been  simulated starting from the discretizations of $L_0$, fixing the value of the bare coupling
and doubling the number of lattice points in each direction.

The simulated quark masses have been halved with respect to the masses simulated on the small volume
in order to have the same amount of the discretization effects proportional to $am$.
The set of parameters for the simulations of this evolution step is reported in \tab{tab:S1simpar}. \\
The step scaling functions at $\beta=6.963$ are plotted, at fixed $m^{RGI}_2$, as 
functions of $m^{RGI}_1$ in {\footnotesize {\bf Figure} [{\bf \ref{fig:A-S1-sigmas}}]}.
As can be seen, $\Sigma(L_0)$ is almost flat in a region of
heavy quark masses starting around the charm mass.
The hypothesis of low sensitivity upon the high--energy scale is thus verified.
The value of the step scaling functions for the $s$ quark are obtained trough linear
interpolation.\\
In {\footnotesize {\bf Figure} [{\bf \ref{fig:A-S1-Cont}}]} are reported the results of the continuum extrapolation
of the 
step scaling function, $\Sigma(L_0)$,
of the pseudoscalar $\overline{h}s$ meson
corresponding to the heaviest quark simulated in this step ($m_h^{RGI} = 3.55$ GeV).\\
The residual heavy mass dependence of the continuum extrapolated step
scaling functions 
is very mild, as shown in {\footnotesize {\bf Figure} [{\bf \ref{fig:A-S1-sigmas2}}]}
in the plot of $\sigma_{B_s}$ as a function of the inverse quark mass. 
The continuum results are linearly extrapolated at the values of the heavy quark masses used
in the small volume simulations.

\noindent The numbers we get at this step are:
\beq
\sigma_{B_s}(L_0) = 0.417(3) 
\qquad \sigma_{D_s}(L_0) = 0.414(3)
\eeq

The step scaling functions  are  free from the systematic errors coming from 
uncertainties on $Z_A$ and $b_A$ since
the multiplicative improvement and renormalization factors cancel exactly in the ratio,  
being the numerator and the denominator evaluated at the same lattice spacing.

\subsection{Second evolution step ($L_1 \to L_2$)}
\label{sec:S2}

\begin{table}[t]
\begin{center}
\scriptsize
\begin{tabular}{cccccc}
\midrule
$\beta$ & $L_1/a$       & $k_c$         & $k$      & $\qquad$ $m^{RGI}$ (GeV) $\qquad$ \\
\toprule
      &                 &               & 0.118128 &  2.012(22)  \\
      &                 &               & 0.121012 &  1.551(17)  \\
      &                 &               & 0.122513 &  1.337(15)  \\
5.960 & $8$             & 0.13490(4)    & 0.131457 &  0.3154(36) \\
      &                 &               & 0.132335 &  0.2322(28) \\
      &                 &               & 0.133226 &  0.1466(44) \\
\midrule
      &                 &               & 0.124090 &  1.984(22)  \\
      &                 &               & 0.126198 &  1.584(17)  \\
      &                 &               & 0.127280 &  1.389(15)  \\
6.211 & $12$            & 0.135831(8)   & 0.133574 &  0.3493(39)  \\
      &                 &               & 0.134177 &  0.2550(29)  \\
      &                 &               & 0.134786 &  0.1510(19) \\
\midrule
      &                 &               & 0.126996 &  1.933(21)  \\
      &                 &               & 0.128646 &  1.547(17) \\
      &                 &               & 0.129487 &  1.355(14) \\
6.420 & $16$            &  0.135734(5)  & 0.134318 &  0.3016(34) \\
      &                 &               & 0.134775 &  0.2038(24) \\
      &                 &               & 0.135235 &  0.1055(15) \\
\bottomrule
\end{tabular}
\mycaption{Simulation parameters for the first evolution step $L_1 \to L_2 = 1.6$ fm.
The RGI quark masses are obtained using eq.~(\ref{eq:rgimass1}).}
\label{tab:S2simpar}
\end{center}
\end{table}

\begin{figure}[t]
\begin{center}
\epsfig{file=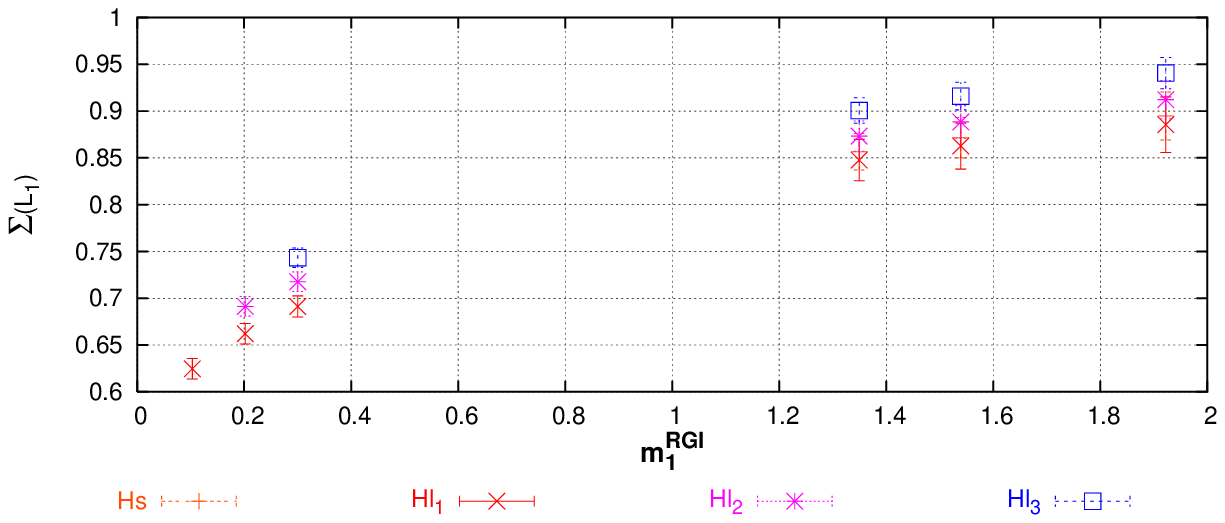,width=12cm}
\mycaption{
The figure shows the step scaling functions $\Sigma(L_1)$ as functions of $m_1^{RGI}$,
for the simulation of the first evolution step
corresponding to $\beta = 6.420$.
The different sets of data correspond to the values of $m_2^{RGI}$.
As can be seen
the step scaling functions approach a plateau for high values of $m_1^{RGI}$. 
Similar plots can be obtained
for the other values of the bare couplings.}
\label{fig:A-S2-sigmas}
\end{center}
\end{figure}

\begin{figure}[t]
\begin{center}
\epsfig{file=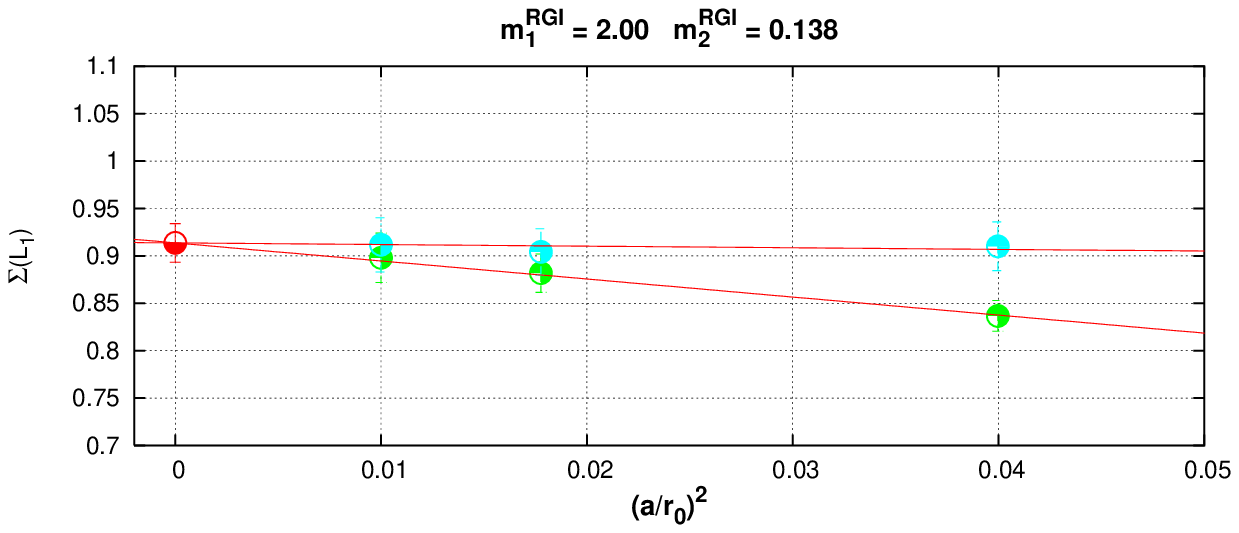,width=12cm}
\mycaption{Continuum extrapolation on the first evolution step, $L_1\mapsto L_2$, of the 
step scaling function, $\Sigma(L_1)$,
of the pseudoscalar meson
corresponding to the heavy quark of mass $m_1^{RGI} = 2.00$ GeV
and to the $s$--quark.
The two sets of data are obtained using the two definitions of RGI quark masses of equations
(\ref{eq:rgimass1}) and (\ref{eq:rgimass2}). 
Units are in GeV.
Similar plots can be obtained for the other 
combinations of quark masses used in our simulations.
\label{fig:A-S2-Cont}}
\end{center}
\end{figure}

\begin{figure}[t]
\begin{center}
\epsfig{file=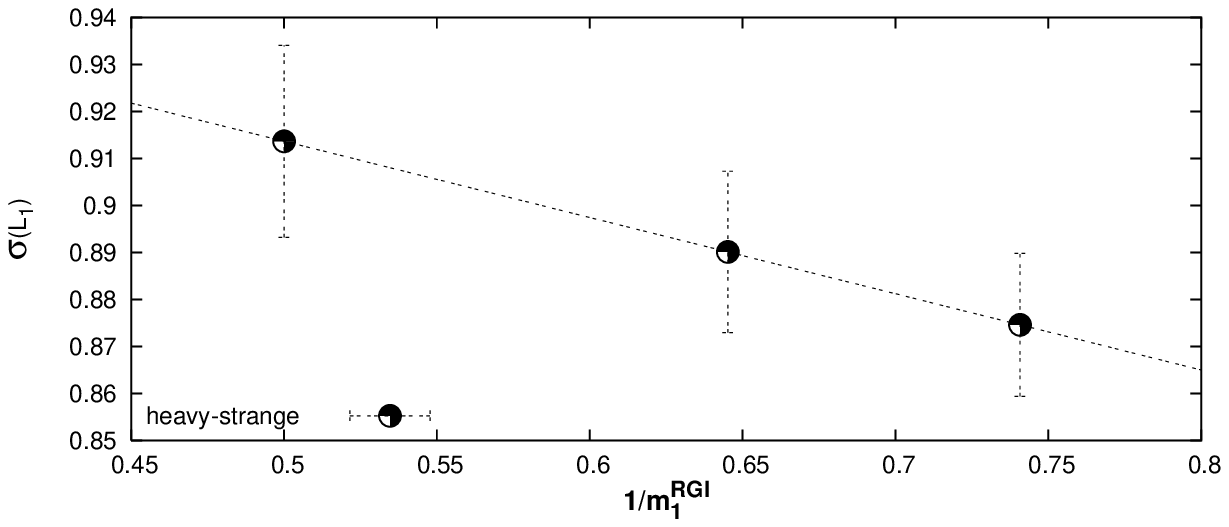,width=12cm}
\mycaption{
The figure shows the continuum extrapolated step scaling functions $\sigma(L_1)$
as functions of $1/m_1^{RGI}$.
The heavy extrapolations are shown only for the $\overline{h}s$ set of data.}
\label{fig:A-S2-sigmas2}
\end{center}
\end{figure}

In order to have the results on a physical volume, $L_2 = 1.6$ fm, a second evolution step
is necessary. This is done computing the step scaling functions
of eq.~(\ref{eq:ssf}) at $L=L_1$, by the procedure outlined in the
previous section. The parameters of the simulations are given in \tab{tab:S2simpar}.

Also in this case, the values of the simulated quark masses have been halved with respect to
the previous step, owing to the lower values of the simulation cutoffs.
Even if we are lowering again the values of the quark masses,
the linear extrapolations at the values of the heavy quark masses
used on the small volume appears to be still valid and under control; see
{\footnotesize {\bf Figure}~[{\bf \ref{fig:A-S2-sigmas},\ref{fig:A-S2-sigmas2}}]}. \\
The value of the step scaling functions for the $s$ quark are obtained trough linear
interpolation. \\
{\footnotesize {\bf Figure} [{\bf \ref{fig:A-S2-Cont}}]} shows
the continuum extrapolation  of the 
step scaling function, $\Sigma(L_1)$,
of the $\overline{h}s$ meson
corresponding to the heaviest quark simulated in this step ($m_h^{RGI} = 2.00$ GeV).

\noindent The numbers for this step are:
\beq
\sigma_{B_s}(L_1) = 0.97(3) 
\qquad \sigma_{D_s}(L_1) = 0.90(2)
\eeq

\section{Physical results}
\label{sec:physresults}

\begin{figure}[t]
\begin{center}
\epsfig{file=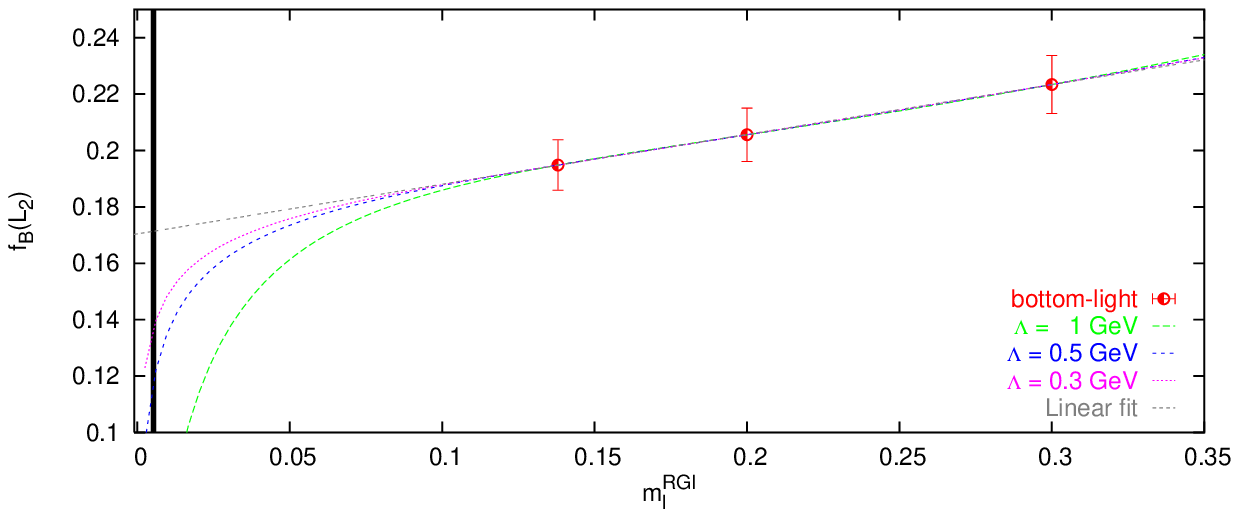,width=12cm}
\mycaption{
Chiral extrapolation of the continuum results for the pseudoscalar 
decay constants on the large volume. Units are in GeV.}
\label{fig:chiral}
\end{center}
\end{figure}

In this section we combine the results of the small
volume with the results of the step scaling functions
to obtain, according to eq.~(\ref{eq:starting}), the physical numbers.
In the end, we get:
\beq
f_{B_s} = 192(6)(4)\ \mbox{MeV} \qquad f_{D_s} = 240(5)(5)\ \mbox{MeV}
\label{eq:fin1}
\eeq
The first error is statistical while the
second one is our estimate of the systematics due
to the uncertainties on the continuum extrapolations, on the
scale and on the renormalization factors, as already
discussed in \sect{SV}. \\
Note that our value for $f_{D_s}$ agrees with the average of dedicated
calculations performed on large volumes \cite{Yamada:2002wh,Becirevic:2002zp}.
This validates our choice of stopping at $L=1.6$ fm that, of course,
can be explicitly checked to be safe by performing further
evolution steps. \\
Using the strategy outlined in the previous sections, 
we have calculated also the decay constant
of the $B_c$ meson. The result we obtain is
\beq
f_{B_c} = 347(5)(8)\ \mbox{MeV}
\label{eq:fbc}
\eeq
that represents the first determination of this quantity
from quenched lattice QCD.

The chiral behavior of heavy--light pseudoscalar decay constants
has been shown \cite{Bernard:1992mk,Sharpe:1992ft,Booth:1995hx,Sharpe:1996qp} 
to contain logarithmic terms  ($\chi$--logs) that are diverging
in the chiral limit, at variance with the unquenched case where these
terms only affect the form of the extrapolation. \\
In {\footnotesize {\bf Figure} [{\bf \ref{fig:chiral}}]} we show
the chiral extrapolations for the continuum heavy--light
pseudoscalar decay constants.
The data corresponding to the different values of the
parameter $\Lambda$ have been extrapolated using the
parametrization suggested in \cite{Sharpe:1996qp}. 
As can be seen from the figure, the presence of
the unphysical quenched $\chi$--logs make the
extrapolations unreliable down to the $u$--quark mass
while the strange region seems to be dominated by a
linear behavior.
Nevertheless, in the literature values
extrapolated linearly in the light quark mass have been quoted. \\
For a \emph{historical} comparison we can quote our own:
\beqa
&&f_B^{linear} = 171(8)(4)\ \mbox{MeV} \qquad f_D^{linear} = 221(7)(5)\ \mbox{MeV}  
\nonumber \\\nonumber \\
&&\frac{f_{B_s}}{f_B^{linear}} = 1.12(2)(1) \qquad \frac{f_{D_s}}{f_D^{linear}} = 1.09(1)(1)
\label{eq:fin2}
\eeqa
that differ by a large factor from the unreliable values obtained
from the fits shown in {\footnotesize {\bf Figure} [{\bf \ref{fig:chiral}}]}
because of the diverging $\chi$--logs.
The numbers quoted above should then be retained for an \emph{historical}
comparison only and should \emph{not} be quoted as the results of
the quenched approximation.

\section{Conclusions}
\label{sec:concl}

In this work we have calculated the $B_s$--meson decay constant
in the continuum limit of quenched lattice QCD.
The results are obtained through a finite volume recursive procedure
where the heavy quark masses have been obtained, 
using the same method, in a previous work.

The main achievement of this computation with respect to
our previous determination of the same quantities are 
the extrapolations to the continuum limit. 
Our systematic errors are due to the extrapolations
to the continuum
and to the physical heavy quark masses.
An additional unknown systematics comes from
the quenched approximation that is believed to produce
visible effects on the meson decay constants.
The known pathological behavior of quenched $\chi$--QCD
does not allow us to quote a quenched value for $f_B$.

Upcoming new powerful super-computers will make
affordable straight calculations of lattice $b$--physics without
recursive methods
but \emph{still} in the quenched approximation.
In this scenario our method could provide the opportunity
of studying \emph{un}quenched $b$--physics 
and/or to deal with other demanding two--scales problems.

\begin{ack}
We want to warmly thank R.~Sommer for a critical reading of the manuscript
and for useful discussions.
This work has been partially supported by the European Community 
under the grant HPRN--CT--2000--00145 Hadrons/Lattice QCD. 
\end{ack}

\bibliographystyle{h-elsevier} 
\bibliography{fbcont}

\begin{thebibliography}{10}

\bibitem{Ciuchini:2000de}
M. Ciuchini et~al.,
\newblock JHEP 07 (2001) 013, hep-ph/0012308.

\bibitem{Hocker:2001xe}
A. Hocker et~al.,
\newblock Eur. Phys. J. C21 (2001) 225, hep-ph/0104062.

\bibitem{Buras:2002yj}
A.J. Buras, F. Parodi and A. Stocchi,
\newblock (2002), hep-ph/0207101.

\bibitem{Stocchi:2002yi}
A. Stocchi,
\newblock (2002), hep-ph/0211245.

\bibitem{Broadhurst:1992fc}
D.J. Broadhurst and A.G. Grozin,
\newblock Phys. Lett. B274 (1992) 421, hep-ph/9908363.

\bibitem{Bagan:1992sg}
E. Bagan et~al.,
\newblock Phys. Lett. B278 (1992) 457.

\bibitem{Neubert:1992sp}
M. Neubert,
\newblock Phys. Rev. D45 (1992) 2451.

\bibitem{Aliev:1983ra}
T.M. Aliev and V.L. Eletsky,
\newblock Sov. J. Nucl. Phys. 38 (1983) 936.

\bibitem{Dominguez:1987ea}
C.A. Dominguez and N. Paver,
\newblock Phys. Lett. 197B (1987) 423.

\bibitem{Narison:1987qc}
S. Narison,
\newblock Phys. Lett. B198 (1987) 104.

\bibitem{Reinders:1988vz}
L.J. Reinders,
\newblock Phys. Rev. D38 (1988) 947.

\bibitem{Colangelo:1991ug}
P. Colangelo et~al.,
\newblock Phys. Lett. B269 (1991) 201.

\bibitem{Penin:2001ux}
A.A. Penin and M. Steinhauser,
\newblock Phys. Rev. D65 (2002) 054006, hep-ph/0108110.

\bibitem{Jamin:2001fw}
M. Jamin and B.O. Lange,
\newblock Phys. Rev. D65 (2002) 056005, hep-ph/0108135.

\bibitem{Yamada:2002wh}
N. Yamada,
\newblock (2002), hep-lat/0210035.

\bibitem{Becirevic:2002zp}
D. Becirevic,
\newblock (2002), hep-ph/0211340.

\bibitem{Becirevic:1998ua}
D. Becirevic et~al.,
\newblock Phys. Rev. D60 (1999) 074501, hep-lat/9811003.

\bibitem{Bernard:1998xi}
C.W. Bernard et~al.,
\newblock Phys. Rev. Lett. 81 (1998) 4812, hep-ph/9806412.

\bibitem{AliKhan:2000eg}
CP-PACS, A. Ali~Khan et~al.,
\newblock Phys. Rev. D64 (2001) 034505, hep-lat/0010009.

\bibitem{Bowler:2000xw}
UKQCD, K.C. Bowler et~al.,
\newblock Nucl. Phys. B619 (2001) 507, hep-lat/0007020.

\bibitem{Lellouch:2000tw}
UKQCD, L. Lellouch and C.J.D. Lin,
\newblock Phys. Rev. D64 (2001) 094501, hep-ph/0011086.

\bibitem{Bernard:2002pc}
MILC, C. Bernard et~al.,
\newblock Phys. Rev. D66 (2002) 094501, hep-lat/0206016.

\bibitem{Kurth:2000ki}
ALPHA, M. Kurth and R. Sommer,
\newblock Nucl. Phys. B597 (2001) 488, hep-lat/0007002.

\bibitem{Heitger:2003xg}
J. Heitger, M. Kurth and R. Sommer,
\newblock (2003), hep-lat/0302019.

\bibitem{Ishikawa:1999xu}
JLQCD, K.I. Ishikawa et~al.,
\newblock Phys. Rev. D61 (2000) 074501, hep-lat/9905036.

\bibitem{AliKhan:2001jg}
CP-PACS, A. Ali~Khan et~al.,
\newblock Phys. Rev. D64 (2001) 054504, hep-lat/0103020.

\bibitem{El-Khadra:1998hq}
A.X. El-Khadra et~al.,
\newblock Phys. Rev. D58 (1998) 014506, hep-ph/9711426.

\bibitem{Guagnelli:2002jd}
M. Guagnelli et~al.,
\newblock Phys. Lett. B546 (2002) 237, hep-lat/0206023.

\bibitem{deDivitiis:2003iy}
G.M. de~Divitiis et~al.,
\newblock (2003), hep-lat/0305018.

\bibitem{Gasser:1985gg}
J. Gasser and H. Leutwyler,
\newblock Nucl. Phys. B250 (1985) 465.

\bibitem{Capitani:1998mw}
S. Capitani et~al.,
\newblock Nucl. Phys. Proc. Suppl. 63 (1998) 153, hep-lat/9709125.

\bibitem{Guagnelli:1998ud}
ALPHA, M. Guagnelli, R. Sommer and H. Wittig,
\newblock Nucl. Phys. B535 (1998) 389, hep-lat/9806005.

\bibitem{Necco:2001xg}
S. Necco and R. Sommer,
\newblock Nucl. Phys. B622 (2002) 328, hep-lat/0108008.

\bibitem{Guagnelli:2002ia}
M. Guagnelli, R. Petronzio and N. Tantalo,
\newblock Phys. Lett. B548 (2002) 58, hep-lat/0209112.

\bibitem{Luscher:1992an}
M. Luscher et~al.,
\newblock Nucl. Phys. B384 (1992) 168, hep-lat/9207009.

\bibitem{Sint:1994un}
S. Sint,
\newblock Nucl. Phys. B421 (1994) 135, hep-lat/9312079.

\bibitem{Luscher:1994gh}
M. Luscher et~al.,
\newblock Nucl. Phys. B413 (1994) 481, hep-lat/9309005.

\bibitem{Bode:2001jv}
ALPHA, A. Bode et~al.,
\newblock Phys. Lett. B515 (2001) 49, hep-lat/0105003.

\bibitem{Guagnelli:2003hw}
Zeuthen-Rome / ZeRo, M. Guagnelli et~al.,
\newblock (2003), hep-lat/0303012.

\bibitem{Luscher:1997ug}
M. Luscher et~al.,
\newblock Nucl. Phys. B491 (1997) 323, hep-lat/9609035.

\bibitem{Luscher:1997jn}
M. Luscher et~al.,
\newblock Nucl. Phys. B491 (1997) 344, hep-lat/9611015.

\bibitem{Sint:1997jx}
S. Sint and P. Weisz,
\newblock Nucl. Phys. B502 (1997) 251, hep-lat/9704001.

\bibitem{deDivitiis:1998ka}
G.M. de~Divitiis and R. Petronzio,
\newblock Phys. Lett. B419 (1998) 311, hep-lat/9710071.

\bibitem{Guagnelli:2000jw}
ALPHA, M. Guagnelli et~al.,
\newblock Nucl. Phys. B595 (2001) 44, hep-lat/0009021.

\bibitem{Garden:1999fg}
ALPHA, J. Garden et~al.,
\newblock Nucl. Phys. B571 (2000) 237, hep-lat/9906013.

\bibitem{Bernard:1992mk}
C.W. Bernard and M.F.L. Golterman,
\newblock Phys. Rev. D46 (1992) 853, hep-lat/9204007.

\bibitem{Sharpe:1992ft}
S.R. Sharpe,
\newblock Phys. Rev. D46 (1992) 3146, hep-lat/9205020.

\bibitem{Booth:1995hx}
M.J. Booth,
\newblock Phys. Rev. D51 (1995) 2338, hep-ph/9411433.

\bibitem{Sharpe:1996qp}
S.R. Sharpe and Y. Zhang,
\newblock Phys. Rev. D53 (1996) 5125, hep-lat/9510037.

\end{thebibliography}

\end{document}